\documentclass[showpacs,floatfix,twocolumn]{revtex4-1}

\usepackage[T1]{fontenc}
\usepackage[utf8]{inputenc}
\usepackage[english]{babel}
\usepackage{pslatex}
\usepackage{graphicx}
\usepackage{latexsym,amsfonts,amsmath,amssymb}
\usepackage[ps2pdf]{hyperref}
\usepackage{paralist}
\usepackage{natbib}

\hypersetup{                                   
  pdfauthor = {Daniel C. Guariento, J. E. Horvath},
  pdftitle = {Consistency of the mass variation formula for black holes
  accreting cosmological fluids},
  pdfsubject = {General Relativity},
  pdfkeywords = {black holes, accretion, general relativity},
  colorlinks = true
}

\newcommand{\ud}{\ensuremath{\mathrm{d}}}

\begin{document}

\title{Consistency of the mass variation formula for black holes
  accreting cosmological fluids}

\author{Daniel C. Guariento}
\email{carrasco@fma.if.usp.br}

\affiliation{Universidade de S\~{a}o Paulo, Instituto de F\'{\i}sica\\
Rua do Mat\~{a}o, Travessa R, 187, 05508-090 Cidade Universit\'{a}ria,
S\~{a}o Paulo -- SP, Brazil}

\author{J. E. Horvath}
\email{foton@astro.iag.usp.br}

\affiliation{Universidade de S\~{a}o Paulo -- Instituto de
  Astronomia, Geof\'{\i}sica e Ci\^{e}ncias Atmosf\'{e}ricas\\
Rua do Mat\~{a}o, 1226, 05508-090 Cidade Universit\'{a}ria, S\~{a}o
Paulo -- SP, Brazil}

\begin{abstract}

We address the spherical accretion of generic fluids onto black
holes. We show that, if the black hole metric satisfies certain
conditions, in the presence of a test fluid it is possible to derive a
fully relativistic prescription for the black hole mass
variation. Although the resulting equation may seem obvious due to a
form of it appearing as a step in the derivation of the Schwarzschild
metric, this geometrical argument is necessary to fix the added degree
of freedom one gets for allowing the mass to vary with time. This
result has applications on cosmological accretion models and provides
a derivation from first principles to serve as a base to the accretion
equations already in use in the literature.

\end{abstract}

\pacs{04.20.-q, 04.70.-s, 98.80.-k}

\maketitle

\section{Introduction}

The accretion of matter onto black holes is one of the cornerstones of
black hole astrophysics. In particular, spherical accretion in a
cosmological context has also been considered \cite{zeldovich-1967}
(see, for instance, the recent review by Carr \cite{carr-2010} and
references therein) and poses interesting problems for the
survival of primordial black holes to the present epoch. Along with
Hawking evaporation, it is essential on providing an accurate and
thorough description for the evolution of the mass of both
astrophysical and primordial black holes throughout the evolution of
the universe.

The most common form of fluid which is considered on accretion models
is a non-self-gravitating (test) perfect fluid, which can be safely
used as a model for most types of ordinary and exotic matter, such as
radiation, cold dark matter and dark energy. Several accretion models
have been proposed for different types of fluids, and their connection
with cosmological models is still an open problem.

The simplest accretion model is the classical Bondi model
\cite{bondi-1951}. Being based on the non-relativistic continuity
equation, it accurately describes accretion onto extense objects, such
as stars. However, due to its Newtonian derivation, its accuracy is
unclear when one considers small distances to a black hole event
horizon, even within the simplest Schwarzschild solution. Another
difficulty arises when using Bondi accretion of a fluid originated
from a scalar field on a general relativistic scenario: it leads us to
consider only the kinetic part of the field
\cite{bean-2002,akhoury-2009,akhoury-2011}. In the general case, the
result is different when a fully relativistic approach is adopted for
the accretion model.

This model can be refined if we consider the relativistic capture of
particles by the black hole \cite{zeldovich-vol1}. Essentially a
gravitational scattering problem, it is most effective in describing
the accretion of radiation \cite{custodio-2002} by the use of a
capture cross-section. However, this model, as well as the Bondi
model, leads to problems when one considers a pressureless fluid
\cite{harada-2005}.

A description of the fluid behavior outside the black hole horizon
without the necessity of considering the individual particles of the
field has been provided by Michel \cite{michel-1972}, based on the
covariant conservation of the fluid's energy-momentum tensor. It links
the fluid evolution with the metric, allowing for a fully relativistic
description of the fluid behavior. However, the Michel model says
nothing about the actual energy transfer into the black hole, whose
description remains left to the Bondi model.

A possible missing piece for the fully relativistic description of
this accretion scenario has been provided in the work by Babichev
\emph{et al.} \cite{babichev-2005} and has since been used in the
literature
\cite{guariento-2007,moruno-2008,sun-2008,jamil-2009,ademir-app-2010,ademir-guariento-2009},
through the use of a mass variation equation of the form

\begin{equation}\label{mponto-babichev}
\frac{\ud m}{\ud t} = \mathcal{A} T_0^{\phantom{0}1}.
\end{equation}

\noindent
where $\mathcal{A}$ is the area of the event horizon and
$T_0^{\phantom{0}1}$ is the $(t,r)$ component of the mixed
energy-momentum tensor of the accreted fluid. Although its seemingly
obvious meaning, equation \eqref{mponto-babichev}, as it is presented
here, can be interpreted in two ways: \begin{inparaenum}[\bfseries
  (a)] \item As a step towards the derivation of the Schwarzschild
  metric, coming directly from the Einstein equations for a
  spherically symmetric space-time \cite{landau-fields}. In that
  specific case, it can be directly obtained from the fluid's
  energy-momentum conservation. \item As a first
  approximation to a fully relativistic continuity equation.
\end{inparaenum}

Both these interpretations suffer from conceptual problems. Assuming
that equation \eqref{mponto-babichev} is obtained from the Einstein
equations, then it would introduce an inconsistency with the
conditions for the derivation of the metric itself. The Schwarzschild
metric is static by construction, which would require the term
$\dot{m}$ to be identically zero. If we abandon this hypothesis, it
would be necessary for consistency to also abandon the static metric
and vacuum hypotheses and perform a full back-reaction analysis
\cite{gao-2008,gonzalez-2009}. This is why the so-called quasi-static
approximation is generally adopted, and the mass evolution is assumed
to be sufficiently slow for the metric to be considered static at each
instant, with a slowly evolving black hole mass function.

Conversely, the continuity equation interpretation is also
insatisfactory, as it lacks a proper rigorous derivation from first
principles and fundamental properties of the energy-momentum tensor.

In this work we provide a simple and exact geometric derivation of
equation \eqref{mponto-babichev} and establish its range of validity,
based on the properties of a space-time with a black hole under
certain assumptions and a generic non-self-gravitating test fluid. The
advantages of achieving such a formal derivation include not only a
better understanding of its origins but it also provides one with the
intuition from a simple case when treating other, more complicated
scenarios.

Here we also adopt the quasi-static hypothesis, and we make no
allusion to the fied equations during the derivations. This decision
is based on the fact that we understand that the black hole metric as
derived from the \emph{vacuum} field equations is now a background for the
accretion process of a \emph{test fluid}, and should take place from
there without any further source terms. Another way of seeing that the
the Einstein equations should not be used during the derivation of
quasi-static accretion is the fact that, once the metric is fixed to
find the vacuum solution, we have no remaining constraints from
the field equations to fix the degree of freedom in $m$. Therefore,
the only self-consistent way to use the field equations to assess
accretion would be through a full back-reaction analysis.

\section{Energy flow through hyper-surfaces}

Let $\ud^3 \Sigma_{\nu}$ be an oriented hyper-surface volume element,
parametrized by coordinates $a$, $b$ and $c$
\cite{misner-thorne-wheeler}. This element may be cast in the
space-time coordinates as
\begin{equation}\label{hiperficie}
\ud^3 \Sigma_{\nu} = \varepsilon_{\nu\alpha\beta\gamma} \frac{\partial
  x^{\alpha}}{\partial a} \frac{\partial x^{\beta}}{\partial b}
\frac{\partial x^{\gamma}}{\partial c} \,\ud a \,\ud b \,\ud c.
\end{equation}

The four-momentum $p^{\mu}$ contained inside such a volume is
given by the value of the energy-momentum which crosses the volume
from past to future at a certain event. Then
\begin{equation}\label{momento-geral}
p^{\mu} = \int_V T^{\mu\nu} \,\ud^3 \Sigma_{\nu}.
\end{equation}

If we take $V$ to be the volume of the object and $u^{\mu}$ its
four-velocity in its rest frame, then integrating equation
\eqref{hiperficie} yields the oriented volume in this frame,
$\Sigma_{\nu} = V u_{\nu}$, which allows us to write equation
\eqref{momento-geral} as
\begin{equation}
p^{\mu} = V T^{\mu\nu} u_{\nu}.
\end{equation}

The energy contained inside the volume $V$, measured in the object's
rest frame, is the four-momentum projection onto the four-velocity
\begin{equation}\label{energia}
E = V T^{\mu\nu} u_{\mu} u_{\nu}.
\end{equation}

\begin{figure}[!htp]
\centering
\includegraphics[width=.45\textwidth]{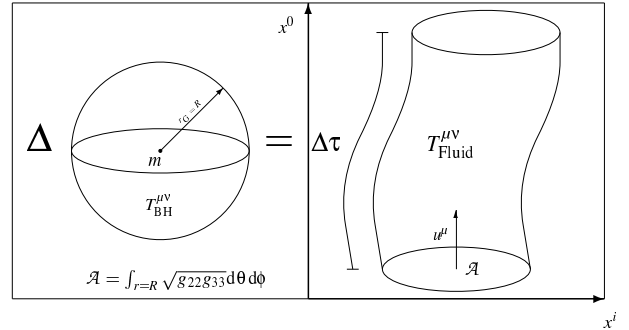}
\caption{Four-momentum variation on the volume inside the event
  horizon of a spherically symmetric black hole by the flux of
  energy-momentum through the boundary of area $\mathcal{A}$. We refer
to $T^{\mu\nu}_{\mathrm{Fluid}}$ in the text simply as $T^{\mu\nu}$.}
\label{magic}
\end{figure}

If there is a flux of material through a limiting surface of the object
during a proper time interval $\Delta \tau$, the four-mo\-men\-tum
variation is given by the value of the energy-momentum which crosses
the volume of another hyper-surface: the world-volume of the
limiting surface during the interval $\Delta \tau$. We illustrate an
example of the construction of this hyper-surface on a
spherically symmetric spacetime with an event horizon located
at $r_{\mathrm{G}} = R$ on figure \ref{magic}.

Thus, the oriented volume of the world sheet given by the surface of
area $\mathcal{A}$ during a proper time interval $\Delta \tau$ is
given by equation \eqref{hiperficie}
\begin{equation}\label{superficie-fluxo}
\Sigma^{\mathrm{BH}}_{\nu} = \mathcal{A} \Delta \tau \sigma_{\nu}
\end{equation}

\noindent
in which $\sigma_{\nu}$ is a unit four-vector orthogonal to the world
sheet. This orientation is a result of the product between the
Levi-Civita symbol $\varepsilon_{\nu\alpha\beta\gamma}$ and the
one-forms tangent to the hyper-surface.

The variation of the object's four-momentum is then
\begin{equation}
\Delta p^{\mu} = \mathcal{A} \Delta \tau T^{\mu\nu} \sigma_{\nu}
\end{equation}

\noindent
and the energy variation measured in the object's rest frame is, as in
\eqref{energia},
\begin{equation}\label{delta-energia}
\Delta E = \mathcal{A} \Delta \tau T^{\mu\nu} u_{\mu} \sigma_{\nu}.
\end{equation}

\section{Energy-momentum inside the event horizon}

The results up to this point are completely general. We now move on to
the specific case when the energy-momentum contained in the interior
region may be identified with the classical energy. This is true on
asymptotically flat space-times.

Let us now assume that the exterior region is filled with a generic
test fluid which flows along the radial direction towards the black
hole interior through the event horizon. This allows us to define the
volume \eqref{superficie-fluxo} as the area of the event horizon
$\mathcal{A}$ times a proper time interval $\ud \tau$. When one moves
sufficiently far from the back hole, there can be associated with the
system an energy, whose variation \eqref{delta-energia} is measured as
\begin{equation}\label{dmdtau-gen}
\frac{\ud E}{\ud \tau} = \mathcal{A} T^{\mu\nu} u_{\mu} \sigma_{\nu}.
\end{equation}

A central point-like object of mass $m$ in empty asymptotically flat
space, as for example a Schwarzschild or Reissner-Nordstr\"{o}m black
hole, may be assigned an energy equal to the ADM mass \cite{adm-1962},
defined as the value of the Hamiltonian for a particular solution of
the field equations at spatial infinity.
\begin{equation}\label{adm}
  E = m \equiv \frac{1}{16 \pi} \oint_S g^{\mu \nu} \left( g_{\mu \alpha, \nu}
    - g_{\mu \nu, \alpha} \right) \ud S^{\alpha}
\end{equation}

\noindent
where $S$ is a 2-sphere with a radially oriented unit length area
element $\ud S^{\alpha}$. The integral must be taken in the limit
where $S$ approaches spatial infinity, and is only defined if the
metric is asymptotically Euclidean \cite{brewin-2007}.

By equating the energy variation inside the horizon to the energy
variation due to the matter flux through the horizon surface, equation
\eqref{dmdtau-gen} becomes
\begin{equation}\label{mdottau}
\frac{\ud m}{\ud \tau} = \mathcal{A} T^{\mu\nu} u_{\mu} \sigma_{\nu}.
\end{equation}

We assume now that the line element with which we are working is
diagonal, as for example is the case of the static Schwarzschild
metric in both isotropic and curvature coordinates, and of the
non-static Vaidya \cite{vaidya-1951-pr} metric, as well as other
solutions of physical interest
\cite{plebanski-1967,delgaty-1998,lake-2003}. We also assume that the
central mass is a function only of time, to avoid arbitrarities when
defining a metric with an extense central object
\cite{vaidya-1951}. Thus, we may cast the left-hand side of equation
\eqref{mdottau} in the rest frame as
\begin{equation}
\frac{\ud m}{\ud \tau} = \frac{\ud m}{\ud t} \frac{\ud t}{\ud \tau} =
\frac{\ud m}{\ud t} \frac{1}{\sqrt{g_{00}}}.
\end{equation}

Writing explicitly $\sigma_{\mu} = (0,\,1,\,0,\,0)$ and using the
four-velocity normalization in the rest frame $u^{\mu} u_{\mu} = 1$,
whose result is
\begin{equation}
u^{\mu} = \left( \frac{1}{\sqrt{g_{00}}},\,0,\,0,\,0 \right) \; ;\quad
u_{\mu} = \left( \sqrt{g_{00}},\,0,\,0,\,0 \right).
\end{equation}

\noindent
we may then rewrite equation \eqref{mdottau} as
\begin{equation}
\frac{\ud m}{\ud t} = g_{00} \mathcal{A} T^{01}.
\end{equation}

By our diagonal metric assumption, we may finally express the mass
variation as \cite{babichev-2005}
\begin{equation}\tag{\ref{mponto-babichev}}
\frac{\ud m}{\ud t} = \mathcal{A} T_0^{\phantom{0}1}.
\end{equation}

\section{Conclusions}

We have worked out in this paper a simple yet formal expression to
deal with the problem of the accretion of a cosmological fluid onto a
black hole. The formal derivation of equation \eqref{mponto-babichev}
completes the framework set up by Michel \cite{michel-1972} in the
test-fluid approximation of spherically symmetric accretion. The
result justifies and clarifies some previous works and constitutes a
reliable prescription which is fully consistent with the general
relativistic conservation equations and does not introduce any
conflicts with the Einstein equations.

In particular, it is now clear what are the requirements the metric
must fulfill in order for equation \eqref{mponto-babichev} to be
valid:

\begin{itemize}

\item There is a background solution to the field equations which
  constitutes a spherically symmetric black hole;

\item The metric is asymptotically flat, so an ADM mass is well
  defined;

\item The line element is diagonal;

\item The accreted fluid is non self-gravitating;

\item Accretion is quasi-static.

\end{itemize}

Equation \eqref{mponto-babichev} should not be interpreted as
depending on local values of $T_0^{\phantom{0}1}$, which is in general also
a function of the radius, but must instead be evaluated in terms of
global (asymptotic) features of the fluid and the black hole. However,
by itself it does not provide us with enough constraints to determine
those invariant values. To acquire the complete picture of the
evolution, we must couple equation \eqref{mponto-babichev} with the
conservation of the energy-momentum tensor and the four-momentum, as
per the procedure derived by Michel \cite{michel-1972}, through the
equations
\begin{gather}
\label{temini-conserva}
T^{\mu\nu}_{\phantom{\mu\nu};\nu} = 0\\
\label{vtemini-conserva}
u_{\mu}T^{\mu\nu}_{\phantom{\mu\nu};\nu} = 0
\end{gather}

In some cases, a first integral of the latter can be obtained,
eliminating the dependence with the radial coordinate
\cite{babichev-2005}. In the particular case of the Schwarzschild
metric accreting a perfect fluid, the expressions
\eqref{temini-conserva} and \eqref{vtemini-conserva} have been fully
worked out \cite{michel-1972,babichev-2004}, and the simplicity of the
Schwarzschild line element does not require the derivation described
here. In fact, the prescription for $\dot{m}$ consisting of solving
the system formed by equations \eqref{mponto-babichev},
\eqref{temini-conserva} and \eqref{vtemini-conserva} is valid for more
general non-static metrics and can be used for any situation which
satisfies the above requirements. In principle, one might also relax
the requirements of asymptotic flatness and the existence of an ADM
mass, as is the case of the McVittie metric \cite{mcvittie-1933} and
its generalizations \cite{faraoni-2007}, provided there can be
defined an analogue to the black hole energy.

In particular, at least within this class of models, the question of
whether a Schwarzschild black hole shrinks in mass when accreting a
perfect phantom field has been clarified, and we now have presented an
additional argument to this discussion, which has emerged in the
literature due to some authors considering a classical energy transfer
\cite{carr-2010,bean-2002,harada-04-2005,pacheco-2007}. More general
cases, for example in which the phantom behavior results from
viscosity terms and/or heat flows
\cite{faraoni-2007,nojiri-2005-2,capozziello-2006,brevik-2005}, can in
principle be handled safely in this framework.

More complicated fluids may lead to different results, even within
this test fluid approximation. It has been pointed out
\cite{jamil-2009} that the mass increasing or decreasing due to the
accretion process of a perfect fluid is directly related to the
violation of the weak energy condition. However, this may not be true
if non-ideal fluids are considered \cite{gao-2008}. Another
possibility is the modification of equation \eqref{mponto-babichev}
due to quantum effects \cite{nojiri-2004}, which must be further
explored within this framework.

\begin{acknowledgments}

The authors wish to thank C. E. Pellicer, G. I. Depetri and A. M. da
Silva for some helpful discussions. This work has been supported by
Funda\c{c}\~{a}o de Amparo \`{a} Pesquisa do Estado de S\~{a}o Paulo
(FAPESP) and Conselho Nacional de Desenvolvimento Cient\'{i}fico e
Tecnol\'{o}gico (CNPq).

\end{acknowledgments}

\bibliography{referencias}

\end{document}